\documentclass{PoS}
\usepackage{caption}
\usepackage{footnote}

\title{The Galactic R Coronae Borealis Stars and the Final 
He-shell Flash Object V4334\,Sgr (Sakurai's Object): A Comparison}

\ShortTitle{R Coronae Borealis Stars and FF Product} 

\author{\speaker{B. P. Hema} \\
        Indian Institute of Astrophysics, II Block Koramangala, Bangalore 560034, INDIA\\
        E-mail: \email{hema@iiap.res.in}}

\author{Gajendra Pandey\\
        Indian Institute of Astrophysics, II Block Koramangala, Bangalore 560034, INDIA\\
        E-mail: \email{pandey@iiap.res.in}}

\author{David L. Lambert\\
        The W.J. McDonald Observatory, University of Texas at Austin, 
	TX78712, USA \\
         E-mail: \email{dll@astro.as.utexas.edu}}

\abstract{The high resolution optical spectra of H-deficient 
stars, R Coronae Borealis stars and H-deficient carbon stars 
are analyzed by synthesizing the C$_{2}$ Swan bands (0,1), (0,0), 
and (1,0) using our detailed line-list and Uppsala model 
atmosphere, to determine the C-abundances
and the $^{12}$C/$^{13}$C ratios which are potential clues to 
the formation process of these stars. 
The C-abundances derived from C$_{2}$ bands are about the 
same for the adopted models constructed with different carbon 
abundances over the range 8.5 (C/He = 0.1\%) to 10.5 (C/He =
 10\%). The carbon abundances derived from C\,{\sc i} lines are a 
factor of four lower than that  adopted for the model 
atmosphere over the same C/He interval, as reported by 
Asplund et al.: `the carbon problem'.
In principle, the carbon abundances obtained from C$_{2}$ Swan 
bands and that adopted for the model atmosphere can be 
equated for a particular choice of C/He that varies from 
star to star (unlike C\,{\sc i} lines). 
Then, the carbon problem for C$_{2}$ bands is 
eliminated. However, such C/He ratios are in general less 
than those of the extreme helium stars, the seemingly 
natural relatives to the RCB and HdC stars. 
The derived carbon abundances and the 
$^{12}$C/$^{13}$C ratios are discussed in 
light of the double degenerate (DD) and 
the final flash (FF) scenarios.
The carbon abundance and the $^{12}$C/$^{13}$C ratios 
for the FF product, Sakurai's Object is derived. 
The carbon abundance 
in the Sakurai's object is 10 times higher than in the 
RCB star VZ\,Sgr. On an average, the carbon abundance in the 
Sakurai's Object is about 10 to 100 times 
higher than in RCB stars. 
The $^{12}$C/$^{13}$C ratio in Sakurai's Object 
is 3.4, the equilibrium 
value, as expected for FF products.}

\FullConference{XII International Symposium on Nuclei in the Cosmos\\
                 August 5-12, 2012\\
                 Cairns, Australia}

\begin{document}

\section{Introduction}

R Coronae Borealis (RCB) stars are F- and G-type 
supergiants having photometric and spectroscopic peculiarities. 
The photometric peculiarity is that, they exhibit visible light 
variability upto several magnitudes at unpredictable times, and, 
the spectroscopic peculiarities are that, they are H-deficient 
having very weak H-Balmer lines in their spectra than expected
for their spectral class and are rich in carbon. The other two groups 
of H-deficient stars that seem to be  close relatives of RCB 
stars are Extreme Helium (EHe) stars that are warmer 
and H-deficient Carbon (HdC) stars that are cooler 
than RCB stars. The EHe and  HdC stars do not 
show photometric variability. 
The origin and evolution of these stars is not well understood 
because of their chemical peculiarity and rare occurrence. 
There are two scenarios in contention, they are, the 
Double Degenerate (DD) Scenario and the Final Flash (FF) Scenario. 
The DD-scenario involves a merger of a helium white dwarf with
 a carbon - oxygen (C - O) white dwarf (Webbink 1984; Iben \&
 Tutukov 1985). The close white dwarf binary resulting from 
a pair of main-sequence stars, merge due to the loss of angular 
momentum by gravitational wave radiation (Renzini 1979), and the merger is 
inflated to a supergiant dimension. The FF scenario involves 
a single post-Asymptotic Giant Branch (AGB) star, experiencing 
a final He-shell flash. 
In investigating the origin and evolution
of these stars, the  determination and interpretation of surface
chemical composition plays the key role.
The chemical composition of RCB and HdC stars determined 
by Asplund et al. (2000) suggested the DD scenario as their 
origin rather than the FF scenario.  
The discovery of high value of $^{18}$O/$^{16}$O ratios 
in cool RCB and HdC stars 
determined from CO infrared bands (Clayton et al. 2005, 2007; 
Garc{\'{\i}}a-Hern{\'a}ndez et al. 2009, 2010), 
and the high fluorine abundances in RCB stars (Pandey et al. 2008) 
provided the evidences for the DD scenario as the origin of 
RCB/HdC stars.
Further potential clues for the origin of these stars can be 
the carbon abundances 
(log $\epsilon$(C)\footnote{Normalized such that
 log $\sum\mu{_i}\epsilon(i)$ = 12.15, where
 $\mu{_i}$ is the atomic weight of element i.})
and the $^{12}$C/$^{13}$C ratios. Because, the low carbon 
abundances and the high $^{12}$C/$^{13}$C ratios are expected 
for the DD scenario and vice-versa for the FF scenario.
For RCB/HdC stars, the carbon abundances derived 
from C\,{\sc i} lines are a factor of four 
lower than that adopted for the model atmospheres constructed with 
different C/He ratios.  That is, the 
predicted strengths of C\,{\sc i} lines are much stronger  than the observed.  
This has been dubbed as the `carbon-problem' by Asplund et al. (2000). 
To further explore the `carbon problem', we have 
derived the carbon abundances using the C$_{2}$ Swan bands in RCB/HdC 
stars including the final He-shell flash object V4334\,Sgr 
(Sakurai's Object). The $^{12}$C/$^{13}$C ratios were also
determined for these stars.

\begin{figure}[ht]
\begin{minipage}[b]{0.50\linewidth}
\centering
\includegraphics[width=\textwidth]{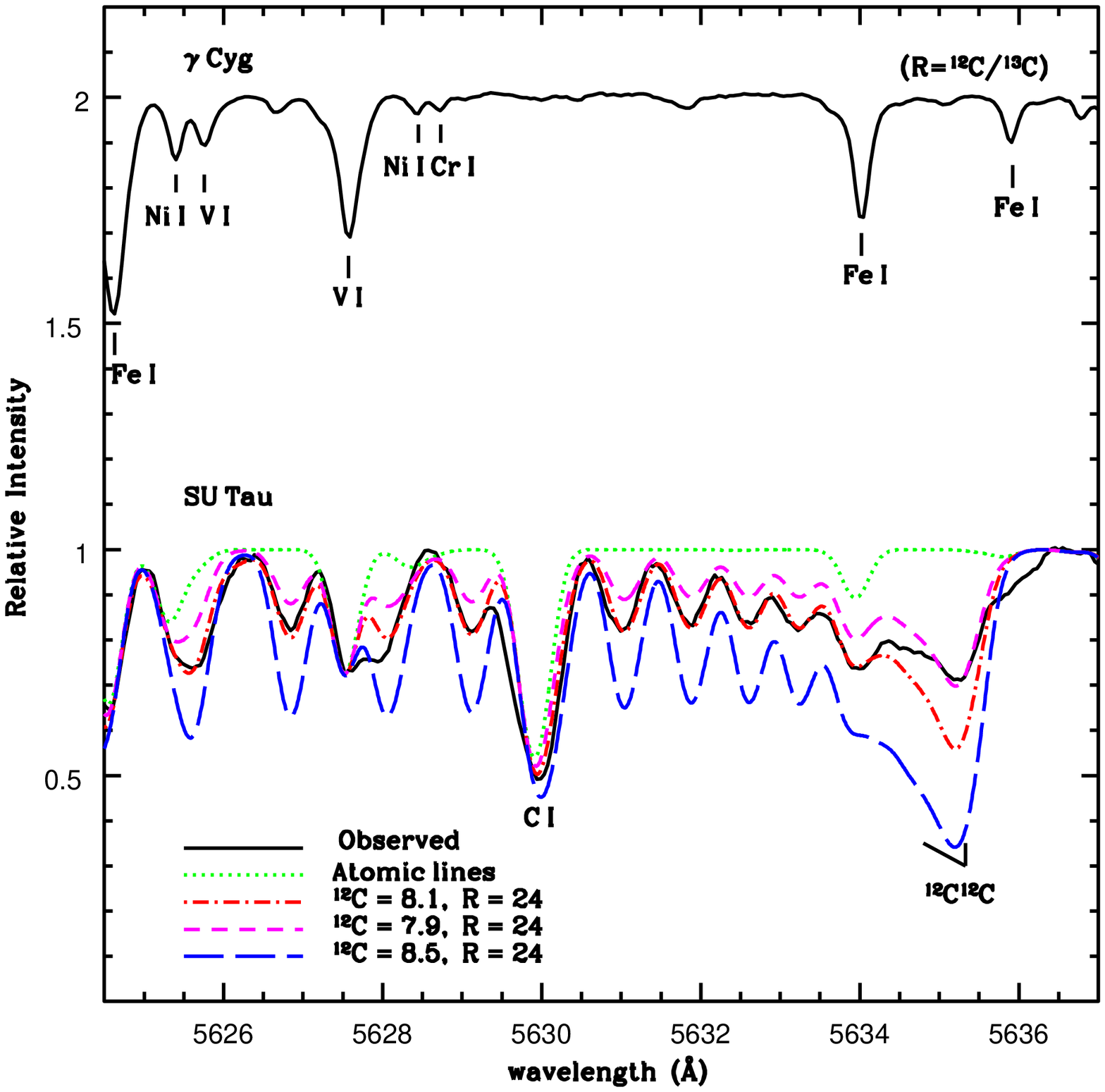}
\caption{Synthesis of the (0,\,1) C$_{2}$ band for SU\,Tau, to determine the 
carbon abundance. The spectrum of the $\gamma$ Cyg is
plotted with the positions of the key lines marked.}
\label{fig:figure1}
\end{minipage}
\hspace{0.6cm}
\begin{minipage}[b]{0.50\linewidth}
\centering
\includegraphics[width=\textwidth]{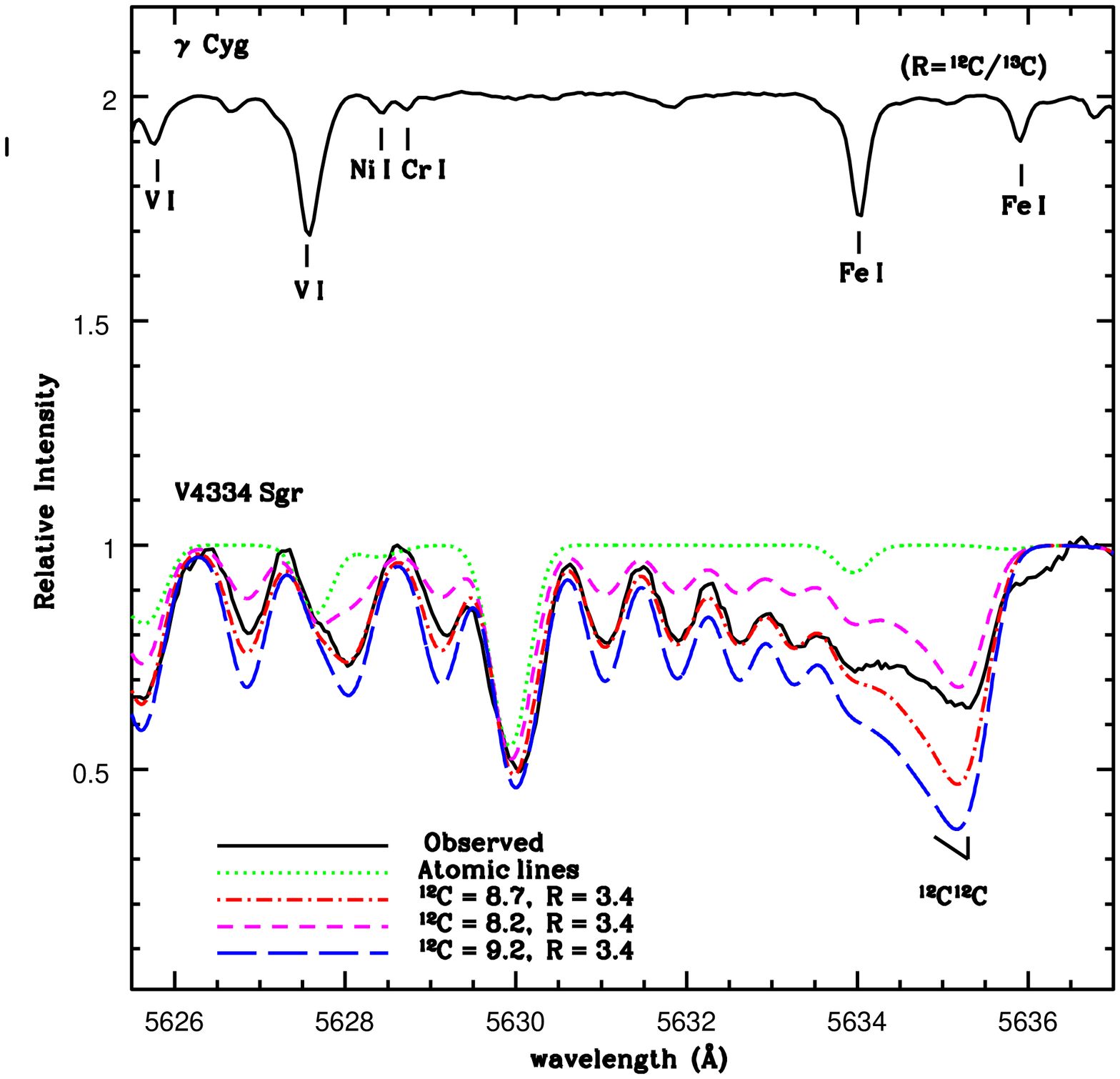}
\caption{Synthesis of the (0,1) C$_{2}$ band for 
V4334\,Sgr (Sakurai's Object) for the stellar parameters 
$T_{\rm eff}$ = 6750 K, log \textit{g} = 0.5, and 
$\xi_{\rm t}$ = 6.5 km\,s$^{-1}$.}
\label{fig:figure2}
\end{minipage}
\end{figure}

\begin{table}
\small
\begin{tabular}{cccccccc}
\hline\hline
Star & &C from (0,1) band & & C from (0,0) band & C from C\,{\sc i} lines\\
& C/He = 0.3\% & C/He = 1\% & C/He = 3\% & C/He = 1\% & C/He = 1\%\\
& log $\epsilon$(C)\,=\,9.0&log $\epsilon$(C)\,=\,9.5&log $\epsilon$(C)\,=\,10.0&log $\epsilon$(C)\,=\,9.5&log $\epsilon$(C)\,=\,9.5\\
\hline
VZ\,Sgr& 9.0 & 8.9 & 8.8 & 8.8 & 8.9\\
UX\,Ant & 8.4 & 8.3 & 8.2 & 8.1 & 8.7\\
RS\,Tel & 8.4 & 8.3 & 8.3 & 8.3 & 8.7 \\
R\,CrB &9.0 & 8.8 & 8.8 & 8.6 & 8.9\\
V2552\,Oph&8.3& 8.1 & 8.2 & 8.1 & 8.7\\
V854\,Cen&8.4& 8.3 & 8.3 & 8.3 & 8.8 \\
V482\,Cyg&8.4 & 8.3 & 8.3 & 8.1 & 8.9\\
SU\,Tau&8.1 & 8.0 & 8.0 & 7.8 & 8.6\\
V\,CrA&8.5& 8.4 & 8.3 & 8.2 & 8.6\\
GU\,Sgr&8.2& 8.1 & 8.1 & 8.1 & 8.9\\
FH\,Sct&7.8 & 7.7 & 7.7 & 7.8 & 8.9\\
V4334\,Sgr$^{1}$ &9.8 &9.7& 9.7 & 9.6& 9.8\\
\hline
\end{tabular}
\begin{footnotesize}
$^{1}$ FF Object
\end{footnotesize}
\caption{The derived carbon abundances for RCB stars and 
FF Product, V4334\,Sgr from (0,\,1) and (0,\,0)
 C$_{2}$ bands and the carbon abundances from the 
C\,{\sc i} lines are also given.}
\label{tab1}
\end{table}

\section{Observations}

The high resolution optical spectra were obtained from McDonald 
Observatory and the Vainu Bappu Observatory. The spectra from 
McDonald Observatory were obtained with the 2.7 m Harlan J. Smith 
telescope and Tull coude cross-dispersed echelle spectrograph 
(Tull et al. 1995). at a resolving power of $\lambda$/d$\lambda$\,=\,60,000. 
The spectra from the Vainu Bappu Observatory were obtained with 
the 2.34 m Vainu Bappu Telescope  equipped with the fibre-fed 
cross-dispersed echelle spectrometer (Rao et al. 2005) and a 
4K$\times$4K CCD are at a resolving power of about 30,000.

\section{Analysis}

The C$_{2}$ Swan bands, $\Delta$$\nu$ = +1, 0, and -1 are 
analyzed for the RCB and HdC 
stars which are having effective temperatures $\leq$ 7000K, as these 
bands are undetectable in warmer stars. 
The (0,0) and (0,1) $^{12}$C$_{2}$ bands 
are used for the determination of the C abundances. 
The (0,1) $^{12}$C$_{2}$ band 
is the best region for the determination of carbon abundance, since it is 
less blended by the atomic lines and  the $^{12}$C$_{2}$ band is free of
 contamination by the $^{12}$C$^{13}$C band, as they are separated by 8\AA\ 
and the $^{12}$C$^{13}$C band is to the blue of the 
blue degraded $^{12}$C$_{2}$ band. 
For deriving the $^{12}$C/$^{13}$C ratio, 
the (1,0)  C$_{2}$ band is used. Because, 
the (1,0) $^{12}$C$^{13}$C band is about 8\AA\ to 
the red of the blue degraded (1,0) 
$^{12}$C$^{12}$C band, the  $^{12}$C$^{13}$C band is
free of contamination by the $^{12}$C$^{12}$C band. 
These C$_{2}$ bands are synthesized using our newly constructed C$_{2}$
molecular line list and including the blending atomic lines for 
all the three regions.  The C$_{2}$ molecular line-list was constructed 
using accurate wavelengths given by Phillips \& Davis (1968) 
and excitation potentials were calculated using the molecular 
constants given by them. The shifts in wavelengths between 
$^{12}$C$^{12}$C and $^{12}$C$^{13}$C lines were calculated 
using standard formulae 
for the vibrational and rotational shifts for C$_{2}$ lines  
(Herzberg \& Phillips 1948; Stawikowski \& Greenstein 1964; 
Russo et al. 2011). The gf-values were calculated from the 
theoretical band oscillator strengths given by Schmidt \& Bacskay (2007). 
The basic data for the atomic lines were taken from NIST and 
Kurucz line list and the gf-values were rederived from the 
spectra of Arcturus (giant), $\gamma$  Cyg (supergiant) and 
Sun (dwarf) and they are in good agreement. For most lines the 
gf-values adopted are those derived from $\gamma$  Cyg spectrum, 
since, $\gamma$  Cyg has the effective temperature and the surface 
gravity that is similar to our program stars.
Using the Uppsala line blanketed H-deficient model atmospheres 
(Asplund et al. 1997a), and the line lists discussed above, 
the C$_{2}$ molecular bands were 
synthesized. The synthesized spectrum was convolved with the 
Gaussian profile with a width that represents effects of stellar 
macroturbulence and instrumental profile, and then matched 
with the observed spectrum.

\section{Results and Discussions}

\subsection{The carbon abundance from C$_2$ bands and the carbon problem} 

The carbon abundance derived from all the three C$_{2}$ bands
are in good agreement. The carbon abundances derived 
from C$_{2}$ Swan bands is about the same
for the adopted models constructed with different carbon abundances
over the range 8.5 (C/He = 0.1\%) to 10.5 (C/He = 10\%) and is different
from the carbon abundances derived from C\,{\sc i} lines. The syntheses of
the C$_{2}$ bands for determining the carbon abundance are shown in figures
1 and 2. 
The derived carbon abundances for RCB stars are given
in Table 1.  
The carbon abundances derived from
C$_{2}$ Swan bands, for the RCB and HdC sample are about 10 times lower
than their close relatives, the EHe stars. This
mismatch, if not a reflection of different modes of formation,
implies that the C abundances for RCB and HdC stars are subject
to a systematic error. Nonetheless, that the carbon abundances
derived from C$_{2}$ Swan bands are the real measure of the carbon
abundances in RCB and HdC stars cannot be ruled out.

\subsubsection{Determination of carbon abundance in FF product:
V4334 Sgr (Sakurai's Object)}

The spectrum of Sakurai's Object obtained from McDonald observatory
on 7$^{th}$ October 1996 is used for our analysis.
The signal-to-noise per pixel calculated in the 4736\AA\ region
is about 140. For the same spectrum, the stellar 
parameters were determined by Asplund et al. (1997b).
These stellar parameters are adopted for our analysis, 
and they are, the effective temperature $T_{\rm eff}$ = 6900$\pm$300 K,  
the surface gravity log \textit{g} = 0.5$\pm$0.3 (cgs units), and 
the microturbulence $\xi_{\rm t}$ = 6.5$\pm$1\,km\,s$^{-1}$.
The RCB star, VZ\,Sgr and Sakurai's Object are
having the near same stellar parameters.
A relative analysis of these two stars is done to derive the 
carbon abundance in Sakurai's Object. The carbon abundance is derived
as discussed in Section 3 and 4.1 (also see Hema et al. 2012).
The carbon abundance in Sakurai's Object is about
0.8 dex higher than in VZ\,Sgr. The carbon 
abundance of VZ\,Sgr is 8.9.
As it is expected, the carbon abundance in the FF product Sakurai's Object is higher 
than the carbon abundance in the RCB stars (see Table 1). 
As an example, the synthesis for Sakurai's Object for deriving the
carbon abundance is shown in Figure 2.

\begin{figure}[ht]
\begin{minipage}[b]{0.50\linewidth}
\centering
\includegraphics[width=\textwidth]{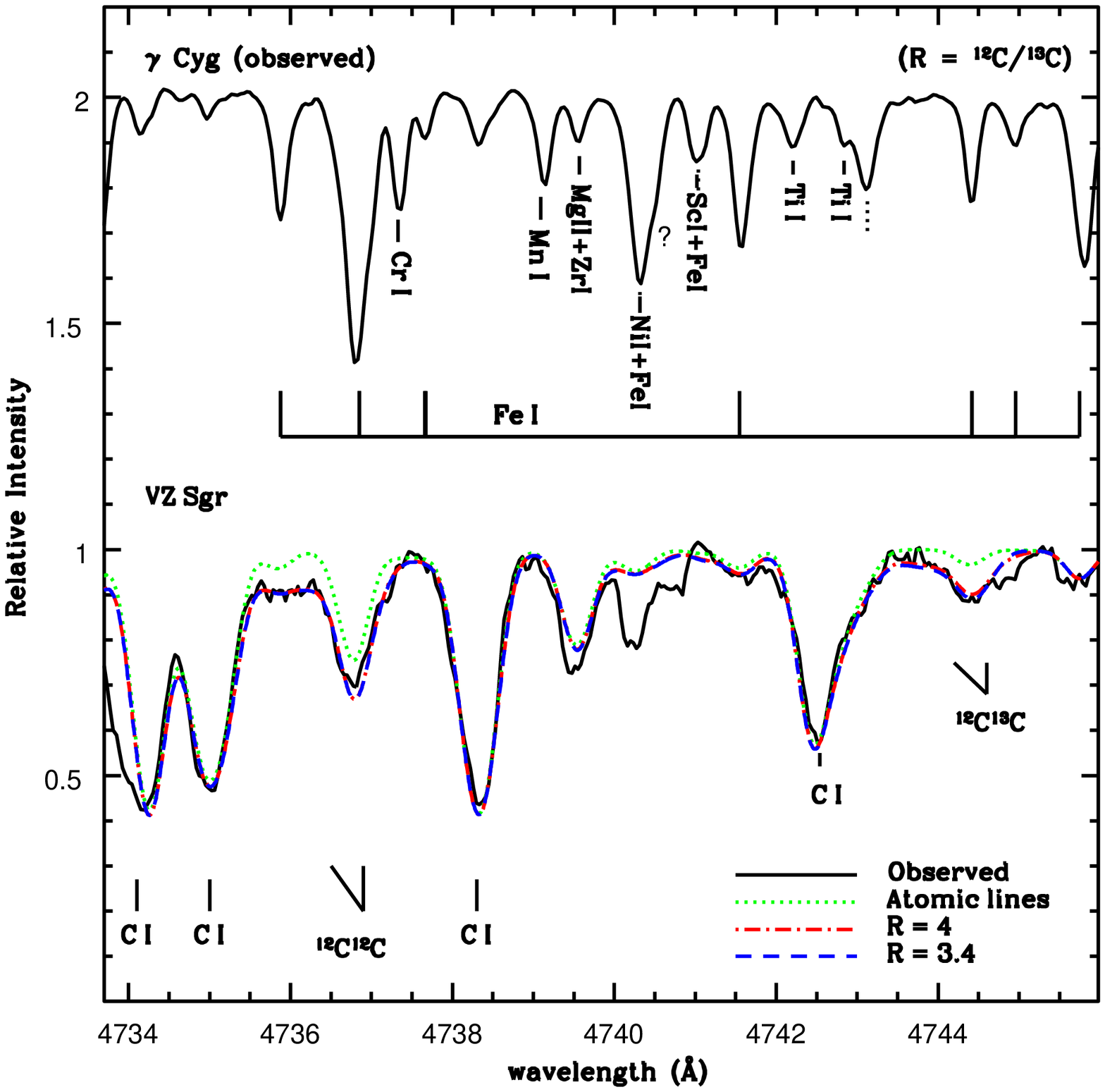}
\caption{Observed and synthetic spectra of the (1,0) 
C$_{2}$ band for VZ\,Sgr, to determine the 
$^{12}$C/$^{13}$C ratio. The spectrum of 
$\gamma$ Cyg is also plotted -- the 
positions of the key lines are also marked.}
\label{fig:figure3}
\end{minipage}
\hspace{0.6cm}
\begin{minipage}[b]{0.50\linewidth}
\centering
\includegraphics[width=\textwidth]{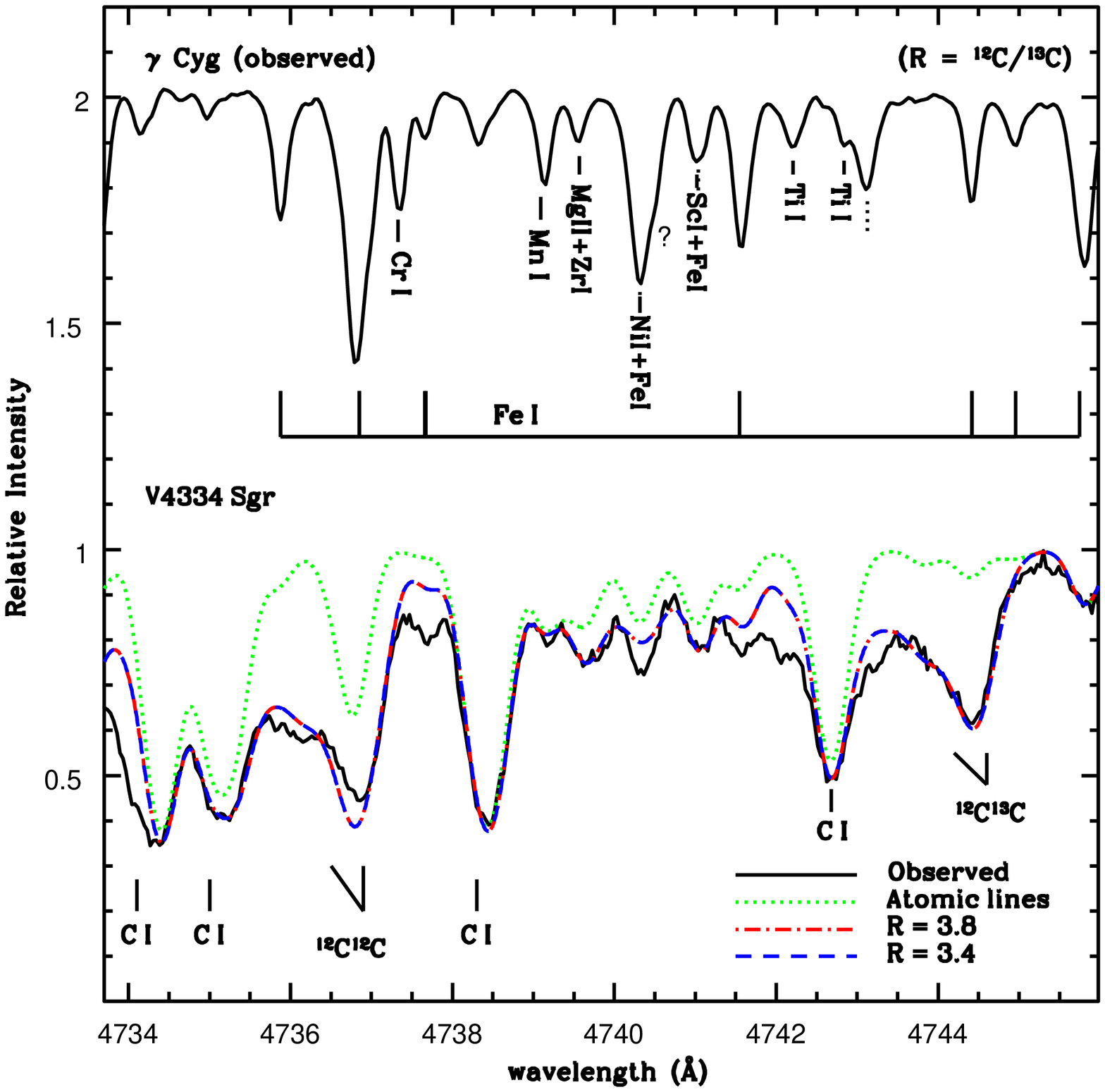}
\caption{Observed and synthetic spectra of the (1,0) C$_{2}$ band for
Sakurai's Object (V4334\,Sgr). The synthesis of the upper limit to the 
$^{12}$C/$^{13}$C ratio (3.8) is also plotted and the
positions of the key lines are also marked.}
\label{fig:figure4}
\end{minipage}
\end{figure}

\subsection{The $^{12}$C/$^{13}$C ratios}

The majority RCB stars and HdC stars are having high values
of $^{12}$C/$^{13}$C ratios and are 
consistent with simple predictions for a
cold  merger of a He white dwarf with a C-O white dwarf.
In DD scenario the carbon abundance is mainly the
triple $\alpha$-processed material $^{12}$C, hence we expect
high values of $^{12}$C/$^{13}$C ratios. 
The minority RCB stars, which
are metal poor and have high [Si/Fe] and [S/Fe] ratios,
relative to majority RCBs: VZ\,Sgr and V\,CrA, are having
low values of $^{12}$C/$^{13}$C ratios (see Figure 3).
The low value of $^{12}$C/$^{13}$C ratio is 
expected for FF scenario, because, with the injection of the surface
protons into the $^{12}$C rich layer, $^{13}$C is synthesized .
Yet, due to their distinctive elemental abundances, 
and the low $^{12}$C/$^{13}$C ratios compared to majority RCB stars,
the origin and evolution of minority RCB stars -- 
VZ\,Sgr and V\,CrA remain unexplained. 
The $^{12}$C/$^{13}$C ratio for Sakurai's Object is derived
using the (1,0) C$_{2}$ band. The  $^{12}$C/$^{13}$C ratio is 
about 3.4, the equilibrium value, as expected for 
the FF product (see Figure 4).

\end{document}